# Elemental depth profiling of fluoridated hydroxyapatite by X-ray photoelectron spectroscopy

Frank Müller [a,*], Christian Zeitz [a], Hubert Mantz [a], Karl-Heinz Ehses [a], Flavio Soldera [b], Matthias Hannig [c], Stefan Hüfner [a], Karin Jacobs [a]

[a] Experimental Physics, [b] Functional Materials,
Faculty of Natural Sciences and Technology, Saarland University, D-66123 Saarbrücken, Germany
[c] Clinic of Operative Dentistry, Periodontology and Preventive Dentistry, Faculty of Medicine - Clinical Medicine, Saarland University Hospital, D-66421 Homburg, Germany

* Corresponding author, Experimental Physics, C 63, Saarland University, 66123 Saarbrücken, Germany. Tel.: +49 681 302 4671; fax: +49 681 302 2947.
E-mail address: f.mueller@mx.uni-saarland.de (Frank Müller)

Structural and chemical changes that arise from a fluoridation of synthetic hydroxyapatite in neutral and acidic fluoridation agents are investigated. For synthetic hydroxyapatite ($Ca_5(PO_4)_3OH$ = HAp), the elemental depth profiles were determined by X-ray photoelectron spectroscopy (XPS) to reveal the effect of fluoridation in nearly neutral (pH = 6.2) and acidic agents (pH = 4.2). Due to the high surface sensitivity of the technique, the depth profiles have a resolution on the nm scale. With respect to the chemical composition and the crystal structure XPS depth profiling revealed very different effects of the two treatments. In both cases, however, the fluoridation affects the surface only on the nm scale which is in contrast to recent literature, where a penetration depth up to several µm was reported. Moreover, evidence is given that the actual fluoridation depth depends on the pH value. Recently, a qualitative three layer model was proposed for the fluoridation of HAp in an acidic agent. In addition to the elemental depth profile, as presented also by various other authors, we present a quantitative depth profile of the compounds $CaF_2$, $Ca(OH)_2$ and fluorapatite (FAp) that are predicted by the three layer model. The analysis of our experimental data exactly reproduces the structural order of the model, however, on a scale that differs by nearly two orders of magnitude from previous predictions. Our results also reveal that the amount of $Ca(OH)_2$ and FAp is small as compared to that of $CaF_2$. Therefore, it has to be questioned whether such narrow $Ca(OH)_2$ and FAp layers really can act as protective layers for the enamel.

## 1. Introduction

The destruction of the enamel by caries is a major oral health problem in industrialized countries. In Germany, for example, the economical damage caused by caries increased from 6.4 billion Euro in 2002 to 7.5 billion Euro in 2004 [1]. Data from numerous clinical investigations clearly demonstrate the cariostatic effect of fluoride compounds in various forms of applications [2]-[5]. Though numerous studies have been performed up to now [6], [7]-[9] no model comprehensively describes the reaction mechanisms of fluoride with enamel. The lack of a model yet renders an optimization of the fluoridating procedure difficult. In a first step, it is useful to establish the chemical and structural effects of the fluoridation. Three main questions have to be answered first: Which chemical species are produced in the enamel by the fluoridation? How deep does the fluoridation reach into the enamel layer? Which, if any, chemical and structural changes in the enamel take place by the fluoridation? One attempt to answer these questions was recently done by Gerth et al. [6], who proposed that, after the fluoridation by an acidic agent, the enamels´ surface can be described by a three-layer-model, containing $CaF_2$, $Ca(OH)_2$, $Ca_5(PO_4)_3F$ and $Ca_5(PO_4)_3OH$. The results of the present work seem to confirm this model, but on a completely different length scale (nm instead of µm). It is the discrepancy of the length scale that motivated the present study, since it questions all earlier interpretations of the effect of fluoride on enamel.
Depth-selective X-ray photoelectron spectroscopy (XPS) is a well-established and well-suited method [10] to study the fluoridation of enamel.

Additionally, scanning force microscopy (SFM) and X-ray diffraction (XRD) will be applied to detect the changes of the surface morphology upon the application of fluoridation. Such studies, however, have been performed with contradictory results, as outlined in. Table I: With respect to the penetration depths of the $F^-$ ions [1], the obtained data do not show a consistent dependence on the agent, the pH value, the fluorine concentration or the exposure time. For example, for a short-term (1 min.) fluoridation of human enamel in an acidic 250 ppm agent (pH = 4), Gerth et al. [6] observed a extended to 5 minutes. The penetration depths reported by Duschner [8] and by Caslavska [9] after the fluoridation of bovine incisors differ only by a factor 2 (2250 nm and 970 nm, respectively), although the exposition times as well as the concentrations and the pH values of the applied NaF agents were strongly different. These two examples suggest that a comparison of data obtained by different authors is questionable and needs to be revisited.

A possible clue to resolve the dissent could lie in the pH of the solution: all studies listed in Table I

| Sample [a] | Agent [b] | pH | concent. | treatment | Decrease length [c] | Penetration Depth [e] | ref. |
|---|---|---|---|---|---|---|---|
| HAp | NaF | 6.2 | 250 ppm | 5min. 37°C | 6 nm | 18 nm | This work |
| BEn | NaF | 6.0 | 1000 ppm | 10min. 37°C | 9 nm (S) [d] <br> 191 nm (B) [d] | 570 nm | [9] |
| BEn | NaF | 6.0 | 30 ppm | 10 min. 37°C | 325 nm [d] | 970 nm | [9] |
| BEn | NH$_4$F | 6.0 | 1000 ppm | 10 min. 37°C | 46 nm [d] | 140 nm | [9] |
| | | | | | | | |
| BEn, HAp | NaF | 5.5 | ~ 22000 ppm (2.3% wt F) | 30 min. 35°C | 12 nm (S) <br> 60 nm (B) | 180 nm | [7] |
| BEn | NaF | 5.0 | ~ 950 ppm (0.1 wt % F) | 30 min. 37°C | 720 nm [d] | 2150 nm | [8] |
| BEn | Amine Hydrofluoride | 5.0 | 0.1 wt % F | 30 min. 37°C | 550 nm [d] | 1650 nm | [8] |
| HAp | OlaFlur | 4.2 | 250 ppm | 5min. 38°C | 1.5 nm (S) <br> 77 nm (B) | 230 nm | This work |
| BEn | NaF | 4.0 | ~ 950 ppm (0.1 wt % F) | 30 min. 37°C | 750 nm [d] | 2250 nm | [8] |
| BEn | Amine Hydrofluoride | 4.0 | 0.1 wt % F | 30 min. 37°C | 1130 nm [d] | 3400 nm | [8] |
| BEn | NH$_4$F | 4.0 | 1000 ppm | 10min. 37°C | 910 nm [d] | 2700 nm | [9] |
| | | | | | | | |
| HEn | OlaFlur | 4.0 | 250 ppm | 1 min. 20°C | 4500 nm | 13500 nm | [6] |

Table I: Comparison of experimental parameters for fluoridation and $F^-$ ions penetration depth as published by various authors. a) HAp: synthetic hydroxyapatite, BEn: bovine enamel, HEn: human enamel. b) For a detailed description of the agent, the reader may consult to the references. c) Some depth profiles cannot be approximated by a single exponential decrease. In these cases, (B) and (S) refer to bulk and surface contributions, d) Extracted from the F intensity ratios for treated and untreated samples. e) Depth at which the F intensity has dropped to 5% of its initial value.

large penetration depth of about 13.5 µm, while in the present study, the fluoridation of a synthetic HAp sample using identical parameters leads to a penetration depth that is about 50 times smaller, namely 230 nm, although the exposition time was agree on an increasing penetration depth with decreasing pH, concurrent with a depletion of phosphorus content. More precisely, the phosphorus content at the surface of a fluoridated sample is lower than that of an untreated sample, whereby the affected depth range of P depletion is similar to that of the F decrease. This indicates a structural change by acidic agents, which may result in less mechanical and chemical stability of the enamel surface.

Recalling literature, fluoridation affects the enamel - consisting of hydroxyapatite crystals

---

[1] The penetration depth is defined as that depth at which the F intensity has dropped to 5% of the initial intensity. Since all depth profiles can be approximated – at least roughly – by an exponential decrease, any other definition of the penetration depth would result in the same relative deviations of the penetration depths.

($Ca_5(PO_4)_3OH$ = HAp) - by replacing $(OH)^-$ groups by $F^-$ ions in HAp, leading to fluorapatite ($Ca_5(PO_4)_3F$ = FAp). This material is more resistant to acids than HAp and is therefore regarded to protect the tooth.

In addition to a pure substitution of $OH^-$ groups, more complex chemical reactions can take place, especially in the acidic range, leading to $CaF_2$ and $Ca(OH)_2$ [6]. Therefore, for a comprehensive understanding of the nature of the fluoridation mechanism, a characterization of the element composition as a function of depth is essential.

In order to yield such results with the highest possible precision a series of experiments was performed under basic, yet thoroughly controlled conditions. For this reason, HAp powder was sintered into a pellet with a density of ~ 90%.

For the fluoridation of the sample, two agents have been applied, one with pH = 4.2 and one with pH = 6.2, in order to be close to the conditions reported in literature (cf. Table I).

Addressing the fundamental problem of a reliable depth scaling, careful attention has been paid to the depth calibration during Ar ion etching. We therefore claim that previous studies reporting depth-resolved elemental composition are very likely off by one or even two orders of magnitude. Thus, in the present study, we will describe method developed to obtain a reliable depth scale in XPS.

## 2. Materials and Methods

### 2.1 Photoelectron spectroscopy

In XPS, slow electrons with kinetic energies in the range of a few hundred eV are used for probing a sample. Due to the small electron mean free path (escape depth ~ 1 - 2 nm cf. Ref. [11]) for electrons in this kinetic energy range, XPS provides a very high surface sensitivity for the analysis of elemental composition. It is a perfect tool to investigate elemental depth profiles, provided the sample can be analysed layer by layer on the nm scale (for details, cf. chapter 4 in Ref. [12]).

Such a nm-scale controlled ablation is possible by Ar ion etching, which is an established tool in surface science [10], [12], [13]. After each etching step, XPS spectra are recorded and the characteristic core level intensities of the respective elements can be quantitatively analyzed [10], [12]. It is therefore of utmost necessity to determine the etching rate with extreme care. By no means it can be inferred solely from the parameters of the Ar ion source (acceleration voltage and emission current), since the real/actual number of Ar ions hitting the surface then remains unchecked (due to instabilities of the ion source, as, e.g., by variations of gas pressure). Instead, the number of Ar ions must be controlled directly via the measurement of the deposited charge on the sample, as will be shown in the following. White light interferometry is then applied to reveal the sputter efficiency with nm resolution in z-scale.

### 2.2 Experimental setup

The experiments were performed with an ESCA Lab Mk II photoelectron spectrometer (by *Vacuum Generators, Hastings, England*) with a ultra high vacuum in the range of $10^{-10}$ mbar. The XPS spectra were recorded with Al-$K_\alpha$ radiation (hν = 1486.6 eV) in normal emission mode. The elemental composition of the samples was determined by using the F-1s, O-1s, Ca-2p, Ca-2s, P-2p and P-2s intensities (normalized to the photon flux) and the corresponding orbital specific photoemission cross section, as listed in Ref. [13]. For Ca and P, the stoichiometry was averaged by the values derived from the 2p and 2s core level intensities.

### 2.3 Preparation of the samples

The preparation of the HAp pellets started with several cycles of filtration of a commercial powder (by *Fluka / SigmaAldrich, Taufkirchen, Germany*) in order to remove impurities such as sodium and chlorine. The powder (~ 1g) was pressed (~8 tons/cm$^2$) to pellets (Ø ~ 16 mm, h ~ 1.8 mm) with green densities of 60-65% (HAp: ρ=3.156g/cm$^3$), and sintered at about 1150°C for several hours with a heating/cooling gradient of about 1°C/min, resulting in densities of 85-90%. The density, given

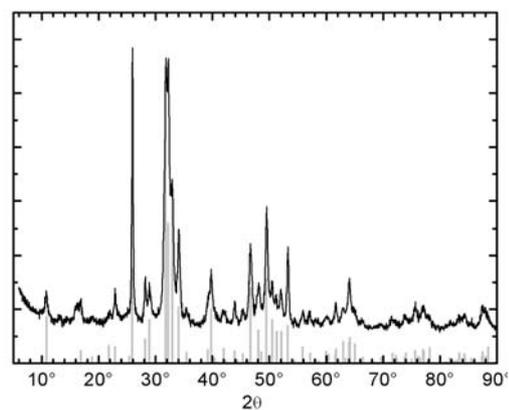

Figure 1: XRD data of an untreated HAp powder, compared to the diffraction pattern of the P63/m space group as taken from Ref. [20].

by the mass-to-volume ratio of a cylindrical pellet, refers to the theoretical density, given by the atomic mass and the volume of a crystallographic a unit cell. A detailed description of the sintering of hydroxyapatite can be found in Ref. [15]. With respect to the chemical composition of the samples, the XPS data provide a stoichiometry of Ca : O : P

= 5 : 12.7 : 3.1, which is very close to the nominal 5 : 13 : 3 ratio of the $Ca_5(PO_4)_3OH$ compound (note that the H content cannot be detected in XPS). With respect to the crystal structure, the XRD data of the powder display the diffraction pattern of the hexagonal $P6_3/m$ space group (cf. Fig. 1).

For fluoridation, the HAp pellets were exposed to a NaF agent (246 ppm, batch 300/053, by *Gaba, Basel, Switzerland*) and an Olaflur© agent (242 ppm, batch 297/038, by *Gaba, Basel, Switzerland*) with pH = 6.2 and 4.2, respectively, 37.5(5) °C for 5 minutes. After the fluoridation, the samples were transferred to the vacuum system in the "wet state" and residues of the agent were removed by evacuating the entrance lock of the vacuum system down to $10^{-5}$-$10^{-6}$ mbar.

**2.4 Calibration of the Ar ion etching dose**

The information needed for a precise depth calibration is the sputter efficiency (etching rate). Therefore, as mentioned before, the achieved etching depth has to be measured as function of deposited Ar ion dose on the sample. The latter can be quantified by measuring the (time-integrated) current that flows off the sample, while the resulting etching depth can be determined by white light interferometry (New View 200 3D Imaging Surface Structure Analyzer by *Zygo Corporation, Middlefield, CT*).

Fig. 2 depicts a schematic drawing of the setup as used for the calibration of the etching rate. Fig. 2a shows the setup of the HAp sample as used during the depth profiling experiments. The HAp sample is covered by a metallic ring that acts as an Ar ion collector, i.e., the Ar ion current (some few μA) that flows off this metallic ring is an instantaneous real time measure for the relative number of Ar ions that hit the sample within each particular period of time:

$$I(Ar^+) \sim \frac{dn^+}{dt}.$$

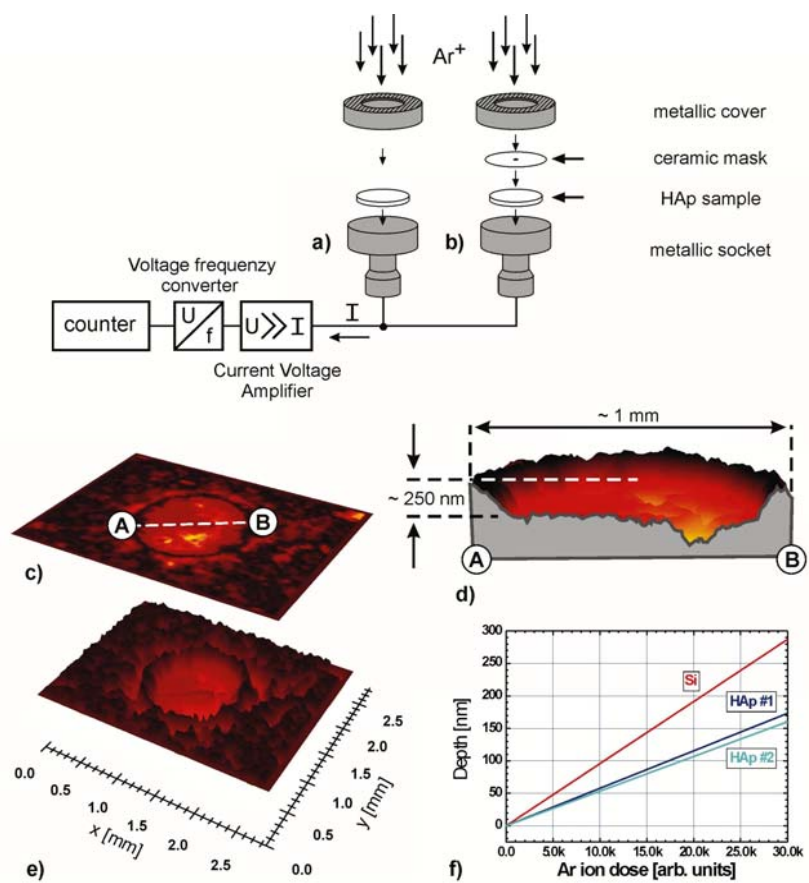

Figure 2: Calibration of Ar etching rate for elemental depth profiling. a) Setup with non-conducting HAp sample during depth profiling. Only the cross-hatched part of the metallic cover contributes to the measured Ar ion current; b) Setup with non-conducting HAp sample for calibration. The non-conducting sample is covered by an also non-conducting mask, so that the same area contributes to the measured ion current as in a); c) Top view of the sputtered hole (after 2 days of Ar ion etching at 4 keV) for a HAp sample as recorded by white light interferometry; d) Depth profile along the A-B direction in c); e) 3D image of c); f) Comparison of the etching rate for a Si and HAp, showing that the etching rate strongly depends on the material.

Since the rate of Ar ion deposition is also a direct measure of the etching rate

$$\frac{dz}{dt} \sim \frac{dn^+}{dt},$$

the total amount of Ar ions that are collected by the metallic ring during a etching period T is then a direct measure for the ablation z of the surface:

$$z = \int_0^T \frac{dz}{dt} dt = c \times \int_0^T \frac{dn^+}{dt} dt = C \times \int_0^T I(Ar^+) dt$$

(with c and C describing constants for the specific experimental arrangement; the determination of C will be described later). The last integral represents the overall charge Q that is "counted" by applying a current-to-voltage amplifier, a voltage-to-frequency converter and a frequency counter, as sketched in Fig. 2 and one has:

$$\int_0^T I(Ar^+) dt \equiv Q$$

(Note: Q includes all fluctuations of the ion current, as, e.g. produced by instabilities of the Ar ion source, and therefore, Q is a measure for the "true" number of Ar ions applied to the sample). Finally, the ablation of the surface for a particular step of Ar ion etching is then given by

$$z = C \times Q.$$

The constant C is determined by the calibration setup, as shown in Fig. 2b. The HAp sample is now additionally covered by a non-conduction ceramic mask with a 1 mm bore hole to ensure that only a small area of the HAp surface is etched. This area is restricted to 1mm in diameter just because of instrumental restrictions by the white light interferometer. At this stage, it is important to note, that the use of a non-conducting ceramic mask guarantees that in both setups, a) and b), the area of the Ar ion collector is the same, namely the area of the metallic ring (the use of a metallic mask during calibration would increase the area of the Ar ion collector, resulting in a large calibration error). For the setup in Fig. 2b, the sample is Ar ion etched for several days (with the same parameters of the Ar ion source as applied during depth profiling) with an overall charge $Q_0$. This procedure results in a hole of 1mm in diameter and a depth $z_0$ of about 250 nm, as determined by white light interferometry, cf. Figs. 2c-e. The constant C for calibration of the ablation depth is then simply given by

$$C = \frac{z_0}{Q_0}.$$

Finally, for the ablation of the HAp surface by applying an arbitrary Ar ion dose Q, one gets:

$$z = \frac{z_0}{Q_0} \times Q$$

Fig. 2f shows, that the etching rate - as expected - strongly depends on the material. While the etching rate for HAp is nearly the same for two different samples (which gives confidence in our method), it strongly deviates from that of a Si wafer. Therefore, an elemental depth profiling by combined XPS/Ar ion etching experiments should always be accompanied by a calibration of the etching rate for the particular sample. Other procedures, as, e.g., referring to manufacturers´ data sheets, would result in large error bars.

## 3. Results

Fig. 3 shows two series of elemental depth profiles of a HAp pellet after fluoridation in a nearly neutral and in an acidic agent. On the ordinate, the elemental composition is given as relative number of atoms. Blue data refer to the NaF agent (246 ppm, pH = 6.2, 5 min. at 37 °C), while the red data refer to the Olaflur© agent (242 ppm, pH = 4.2, 5 min. at 38 °C). There are obviously strong differences between both series.

For the sample exposed to the pH = 6.2 agent (blue circles), only the $F^-$ signal in Fig. 3e shows noticeable depth dependence, while the signals for Ca, O and P in Fig. 3a and Figs. 3c-d stay roughly constant. From the distribution of the C signal in Fig 3b, representing an unavoidable, but non-relevant impurity, it is evident that all "ex situ" effects, as, e.g., adsorbates or residues of the agent, are removed after 2-3 steps of preparation. The Ca, O and P contributions can be approximated by nearly constant values that display the elemental composition of the untreated sample. Hence, fluoridation in a nearly neutral solvent does not affect the initial Ca : P : O ratio. If the $F^-$ distribution in Fig. 3e is approximated by a simple exponential decrease, one finds a decrease length of about 6 nm with an amplitude of about 4% of the total oxygen amount, i.e., the initial $F^-$ concentration corresponds to 50% of the $(OH)^-$ groups (this 4% effect is smaller than the scattering of the data in Fig. 3a). At this stage, it is important to point out that the penetration depth [1] of the F⁻ ions is about 18 nm, a value much smaller than hitherto reported in literature [6]-[9], cf. also Table I.

For the sample exposed to the pH=4.2 solution (red data), considerable differences in the depth profile relative to the pH=6.2 sample can be observed for all elements. First of all, the F⁻ ions in Fig. 3e obviously range much deeper into the sample (far beyond measured here) and the initial F intensity has increased by more than one order of magnitude compared to the value for the NaF agent (note, that in Fig. 3e also the ordinate has an axis break). In addition, however, considerable reductions in the O signal in Fig. 3a as well as in the P signal in Fig. 3d can be observed, while, in contrast, the amount of Ca in Fig. 3c is slightly increased. These different distributions of Ca, O and P demonstrate a considerable change in the chemistry and most likely in the structure of the HAp sample due to the acidic fluoridation, probably to more volatile species. This means that a much deeper fluoridation takes place, which, however, is accompanied by a modification of the crystal structure in the surface region.

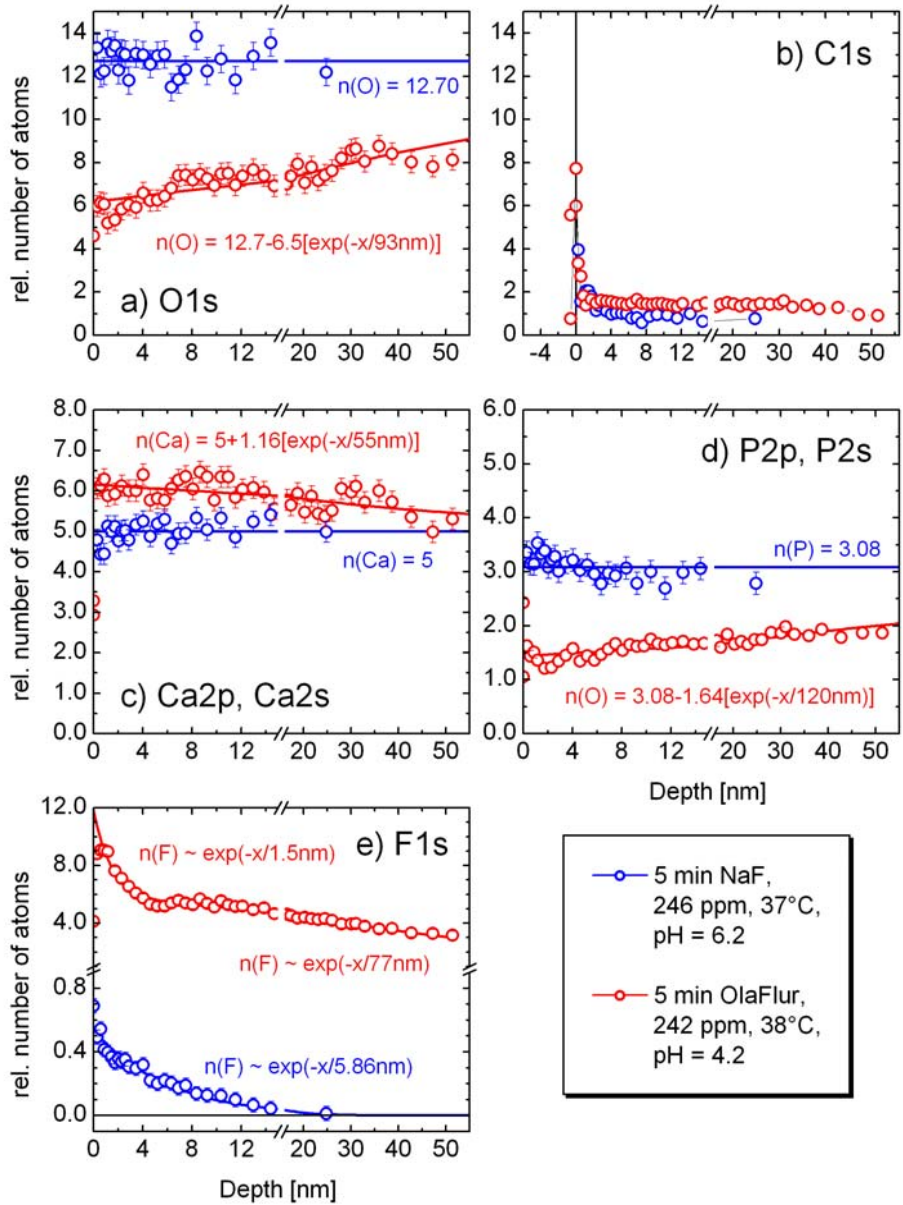

Figure 3: XPS depth profiling (atomic ratios) of a synthetic HAp sample after 5 min. of fluoridation in NaF (246 ppm, pH=6.2, 37 °C, blue circles) and in OlaFlur (242 ppm, pH=4.2, 38°C, red circles). a) Oxygen, b) Carbon, c) Calcium, d) Phosphorus, e) Fluor. Note the break of the abscissa for all elements and the additional break of the ordinate for Fluor.

al. [6], i.e., the surface region is formed by $CaF_2$, followed by $Ca(OH)_2$ and the FAp and HAp apatites. Therefore, the proposed three-layer-model seems to be valid, although the term "layer" seems to be unfavourable, because these "layers" exhibit a strong intercalation instead of forming sharp

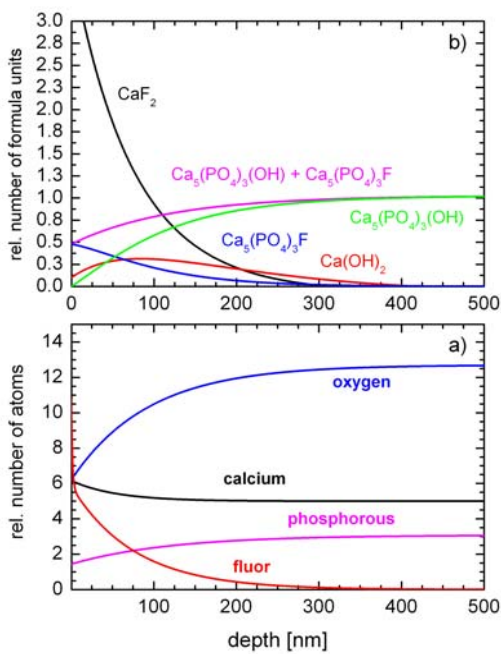

Figure 4: a) Approximation of the elemental depth profiles from Fig. 3 by exponential decay/growth. b) Depth profiles of the compounds $CaF_2$, $Ca(OH)_2$, FAP and HAp, as calculated by the elemental distributions from a).

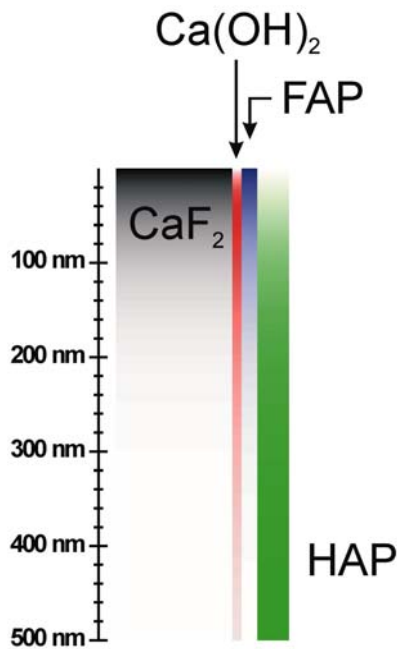

Figure 5: Schematic drawing of the depth profile for the HAp sample after fluoridation in Olaflur®. The linear color scale represents the relative depth distribution of each compound, while the width of

Such structural modifications were supposed in the recent study by Gerth et al. [6]. For the fluoridation within an acidic agent, the authors proposed a three layer model whereby the fluoridated enamel is formed by a stacking (from outside to inside) of $CaF_2$, $Ca(OH)_2$, $Ca_5(PO_4)_3F$ and $Ca_5(PO_4)_3(OH)$. In order to get a quantitative depth profile for these compounds by using the data from Fig. 3, the elemental depth profiles were approximated by exponential decrease/increase in Fig. 4a. We hereby took as a constraint that the limiting values are given by the values of the NaF data (blue) in Fig. 3, which also represent the values for an untreated sample due to the small overall content of F. From the elemental depth profiles, it is straightforward to extract the depth profiles of the compounds by their stoichiometry, as depicted in Fig. 4b. Although the calculation of the depth profiles of the compounds was not restricted by a required order of the compounds, the result exactly displays the configurational order that was proposed by Gerth et interfaces. Within this context, the results also show that the FAp and $Ca(OH)_2$ "layers" provide only a small overall contribution after the surface modification by the fluoridation, compared to the large amount of $CaF_2$. In general, the fluoridated sample is mainly formed by $CaF_2$ and HAp, where $CaF_2$ switches to HAp at about 125 nm, as shown in Fig. 5.

## 4. Discussion

According to the data of Fig. 3, the fluoridation of a HAp sample by a nearly neutral solution leads to the result that the $F^-$ ions penetrate the sample only at a depth of about 18 nm, a value much smaller (up to 1 to 2 orders of magnitude) than reported so far in literature (see Table I). Assuming an annual wear of human enamel of about 1000 nm (1 $\mu$m, [16]), it is realized that the penetration depth of the $F^-$ ions during fluoridation in nearly neutral solution (~ 18 nm 5%-length, ~ 6 nm decrease constant ) is of the order of the daily wear of the teeth (~ 3 nm, cf. Ref. [16]). In agreement with other investigations [7]-[9], no change in the P, Ca and O composition of the HAp sample is found by the nearly neutral fluoridation using NaF.

In contrast, fluoridation in acidic solution (pH = 4.2) considerably changes the P, Ca and O concentration. Depending on the analytical methods used there are reports in the literature on a change in crystal structure to a depth of several $\mu$m [6], [7], [9]. Such a change of crystal structure is also confirmed by our data in Fig. 3 and Fig. 4, but there are strong differences with respect to the depth scale. According to Fig. 4b, the HAp is no longer affected by the fluoridation beyond a depth of about 200-300 nm. Irrespective, whether the $F^-$ ions penetrate into the enamel up to 200 nm (our value) or up to several µm (literature values, cf. Table I), the results also show, that just one application of an acidic agent results in a penetration depth, that approximately amounts to more than the monthly (from our values) or yearly (from literature values) wear of a tooth, thus influencing the surface of the tooth considerably.

So far, it is not entirely clear what kind of surface modification is produced by the pH = 4.2 treatment, but our results from Fig. 3 and Fig. 4 give evidence that the three layer $CaF_2$ - $Ca(OH)_2$ - FAp/HAp model by Gerth et al. [6] is valid. The quantitative analysis of the data also show, however, that the $Ca(OH)_2$ and FAp contributions only play a secondary role. Due to their small overall amount it is therefore questionable if the $Ca(OH)_2$ and FAp contributions can indeed provide antimicrobial and acid-resistant buffer layers, respectively, as proposed by Gerth et al. [6]. The surface area is mainly dominated by large amounts of $CaF_2$, and, as a consequence, such a surface may be less stable than the original one containing HAp. Due to the fluoridation at pH = 4.2, a deep penetration of caries protecting $F^-$ ions is achieved, but according to Fig. 4 and Fig. 5 rather as $CaF_2$ than as the protecting FAp.

It has been shown previously by AFM investigations [17] that the enamel surface is covered by a continuous layer of spherical $CaF_2$ globules (diameter of 120- 300 nm) within minutes after treatment with 0.1% amine fluoride (pH = 4.5), whereas treatment with a neutral NaF solution caused formation of only a few scattered $CaF_2$ globules. These findings indicate that the fluoride deposition in form of $CaF_2$ precipitates is affected by the pH value of the fluoride containing solution. The accelerated dissolution of the enamel apatite at lower pH provides more free calcium ions and, thus, it favours the precipitation of the $CaF_2$ globules. The AFM observations reported by Petzold [17] fit well with the present XPS data, especially with regard to the depth of fluoridation and chemical modification of the hydroxyapatite surface (high amount of $CaF_2$) after application of the acidic amine fluoride solution for 5 min.

Also with respect to a molecular dynamic simulation by de Leeuw [18] of the F-HAp interaction, the small decay length of about 6 nm for nearly neutral agent in Fig. 3e provides an interesting result because this small value seems to be confirmed by the calculated value of about 3 unit cell distances (3 nm), i.e., these calculations predict, that $F^-$ replaces $(OH)^-$ within the three topmost layers of the HAp sample.

Thus, a treatment with nearly neutral solution (pH = 6.2) seems more advisable, because it causes no structural effects at the enamel surface. However, since the F ions have only a small penetration depth, F ions have to be continuously administered (as, e.g., by the use of a neutral fluoridated toothpaste or mouthrinse) since the protecting FAp layer will otherwise be partly removed by daily tooth brushing.

In principle, fluoride can induce cariostatic effects by (1) reducing enamel solubility when incorporated into the mineral structure; (2) by fostering the remineralization of incipient enamel lesions and the deposition of fluoridated phases (within dental plaque) which provide a source of mineral ions (Ca, P, F) under acidic conditions; and (3) by reducing the net rate of transport of matter out of the enamel surface, under acidic conditions, by inducing the reprecipitation of fluoridated hydroxyapatite phases within enamel. It has been concluded by Margolis and Moreno [19] based on an analysis of clinical and laboratory data that the benefits provided by fluoridation result, to a large degree, from topical effects. Thus, clinical procedures should be developed to establish and maintain low levels of free fluoride in plaque fluid. This will require frequent exposure to topical fluorides [19].

Summing up the discussion, the results of the present study on the fluoridation of synthetic HAp samples match qualitatively the results reported in literature. First, the fluoridation by a nearly neutral agent leads to a substitution of the $(OH)^-$ group by the $F^-$ ions with no changes in the chemistry and structure of the initial compound, i.e. a partial HAp to FAp transition takes place. Secondly, the fluoridation by an acidic agent results in considerable changes in the surface structure of HAp by the formation of other compounds, such as $Ca(OH)_2$, FAp and $CaF_2$.

However quantitatively, the results of the present investigation are in contrast to previous studies: First, the values of the penetration depth of F, however, are considerably lower (one to two orders of magnitude!) as compared to the values of other authors, though the fluoridation was performed under similar conditions (cf. Table I). One could argue, however, that the discrepancy arises from using synthetic HAp instead of natural enamel, yet in the case of a nearly neutral agent, our low values of the penetration depth are in accordance to dynamical simulations. It seems to be more likely that such large discrepancies result from a different calibration of depth scales in XPS spectra. Secondly, for the compounds which are formed by fluoridation in an acidic agent the quantitative analysis of the depth profiles shows, that it is rather $CaF_2$ than FAp that is produced by this treatment.

**Acknowledgement**

This work was supported by the *Deutsche Forschungsgemeinschaft* via the *Sonderforschungsbereich 277 "Grenzflächenbestimmte Materialien"*. The authors thank H. P. Beck, R. Haberkorn, H. Kohlmann, U. Hartmann and A. English for providing the setup for the preparation of the HAp pellets.

**References**

[1] Statistisches Bundesamt, Gesundheit und Krankheitskosten 2002 und 2004, Wiesbaden, Germany (2007)

[2] A.R. Ten Cate, *Oral Histology: development, structure, and function*, 5th ed. (1998), St. Louis, C. V. Mosby

[3] M.H. Ross, G.I. Kaye, W. Pawlina, *Histology: a text and atlas*, 4th ed. (2003), Philadelphia; London: Lippincott Williams & Wilkins

[4] H.T. Dean, *Endemic fluorosis and its relation to dental caries*, Public Health Reports **53** (1938), 1443

[5] H.T. Dean, F.A. Arnold, E. Elvone, *Domestic water and dental caries. V. Additional studies of the relation of fluoride domestic waters to dental caries experience in 4425 white children aged 12 to 14 years of 13 cities in 4 states*, Public Health Reports **57** (1942), 1155

[6] H.U.V. Gerth, T. Dammaschke, E. Schäfer, H. Züchner, *A three layer structure model of fluoridated enamel containing $CaF_2$, $Ca(OH)_2$ and FAp*, Dent. Mater. **23** (2007), 1521

[7] H. Uchtmann, H. Duschner, *Electron spectroscopic studies of interactions between superficially applied fluorides and surface enamel*, J. Dent. Res. **61** (1982), 423

[8] H. Duschner, H. Uchtmann, *Effect of sodium fluoride, stannous fluoride, amine hydro fluoride and sodium monofluorophosphate on the formation of precipitates adhering to bovine enamel*, Caries Res. **22** (1988), 65

[9] V. Caslavska, H. Duschner, *Effect of a surface-active cation on fluoride/enamel interactions in vitro*, Caries Res. **25** (1991) 27

[10] H. Oechsner (ed.): *Thin film and depth profile analysis*, Topics in Current Physics **37** (1984), Springer, Berlin, Heidelberg, New York, Tokyo

[11] M.P. Seah, W.A. Dench, *Quantitative electron spectroscopy of surfaces: A standard data base for electron inelastic mean free paths in solids*, Surf. Interface Anal. **1** (1979) 2

[12] D. Briggs, M. P. Seah *Practical Surface Analysis -Auger and X-ray Photoelectron Spectroscopy*, Wiley Interscience, 1990 (2nd ed.)

[13] G. Carter, V. Vishnyakov, *$Ne^+$ and $Ar^+$ Ion Bombardment-induced Topography on Si*, Surf. Interface Anal. **23**, (1995) 514

[14] J.J. Yeh, I.Lindau, *Atomic Data and Nuclear Data Tables*, **32**, 1-155 (1985)

[15] O. Prokopiev, I. Sevostianov, J. Genin, S. Munson McGee, C. Woodward, *Microstructure and elastic properties of sintered hydroxyapatite,* International Journal of Fracture **130**, L183-L190, (2004)

[16] A. Joiner, C. J. Philpotts, T. F. Cox, A. Schwarz, K. Huber, M. Hannig, *The Protective Nature of Pellicle towards Toothpaste Abrasion on Enamel and Dentine,* Journal of Dentistry **36** (2008), 360

[17] M. Petzold, *The influence of different fluoride compounds and treatment conditions on dental enamel: a descriptive in vitro study of the $CaF_2$ precipitation and microstructure*, Caries Res. **35** (2001), 45-51

[18] N.H. de Leeuw, *Resisting the onset of hydroxyapatite dissolution through the incorporation of fluoride,* J. Am. Chem. Soc. B **108** (2004), 1809

[19] H.C. Margolis, E.C. Moreno, *Physicochemical perspectives on the cariostatic mechanisms of systemic and topical fluorides*, J Dent Res. **69** (1990), 606-613

[20] Entry 74-565 of database Powder Diffraction File PDF 2, Release 2001, International Centre for Diffraction Data (ICDD), Newtown Square, Pennsylvania, USA